\documentclass[twocolumn]{aastex61}

\newcommand\aastex{AAS\TeX}

\received{}
\revised{}
\accepted{}
\submitjournal{ApJS}

\shorttitle{\aastex\ Cold condensation in the ISM}
\shortauthors{Fulvio et al.}

\begin{document}

\title{Laboratory experiments on the low temperature formation\\
of carbonaceous grains in the ISM}

\correspondingauthor{Daniele Fulvio}
\email{dfulvio@puc-rio.br, dfu@oact.inaf.it}

\author{Daniele Fulvio}
\affil{Departamento de F\'{\i}sica, Pontif\'{\i}cia Universidade Cat\'{o}lica do Rio de Janeiro, Rua Marqu\^{e}s de S$\tilde{\textrm{a}}$o Vicente 225, 22451-900, Rio de Janeiro, RJ, Brazil}
\affiliation{Laboratory Astrophysics Group of the Max Planck Institute for Astronomy at the Friedrich Schiller University Jena, Institute of Solid State Physics, Helmholtzweg 3, D-07743 Jena, Germany}

\author{S\'{a}ndor G\'{o}bi}
\affiliation{Institute for Geological and Geochemical Research, Research Centre for Astronomy and Earth Sciences, Hungarian Academy of Sciences, 45 Buda\"{o}rsi street, 1112 Budapest, Hungary}
\affiliation{Present address: Department of Chemistry, University of Hawaii at Manoa, 2545 McCarthy Mall, Honolulu, HI 96822, USA}

\author{Cornelia J\"{a}ger}
\affiliation{Laboratory Astrophysics Group of the Max Planck Institute for Astronomy at the Friedrich Schiller University Jena, Institute of Solid State Physics, Helmholtzweg 3, D-07743 Jena, Germany}

\author{\'{A}kos Kereszturi}
\affiliation{Konkoly Thege Mikl\'{o}s Astronomical Institute, Research Centre for Astronomy and Earth Sciences, 1121 Budapest, Konkoly Thege Mikl\'{o}s str. 15-17, Hungary}

\author{Thomas Henning}
\affiliation{Max Planck Institute for Astronomy, K\"{o}nigstuhl 17, 69117 Heidelberg, Germany}


5
\begin{abstract}

The life-cycle of cosmic dust grains is far from being understood and the origin and evolution of interstellar medium (ISM) grains is still under debate. In the ISM, the cosmic dust destruction rate is faster than the production rate by stellar sources. However, observations of ISM refractory matter suggest that to maintain a steady amount of cosmic grains, some supplementary production mechanism takes place. In this context, we aimed to study possible re-formation mechanisms of cosmic grains taking place at low temperature directly in the ISM. The low temperature condensation of carbonaceous materials has been investigated in experiments mimicking the ISM conditions. Gas-phase carbonaceous precursors created by laser ablation of graphite were forced to accrete on cold substrates (T $\approx$  10 K) representing surviving dust grains. The growing and evolution of the condensing carbonaceous precursors have been monitored by MIR and UV spectroscopy under a number of experimental scenarios. It is demonstrated, for the first time, the possibility to form ISM carbonaceous grains ``in situ". The condensation process is governed by carbon chains that first condense into small carbon clusters and finally into more stable carbonaceous materials, which structural characteristics are comparable to the material formed in gas-phase condensation experiments at very high temperature. We also show that the so-formed fullerene-like carbonaceous material is transformed into a more ordered material under VUV processing. The cold condensation mechanisms here discussed can give fundamental clues to fully understand the balance between the timescale for dust injection, destruction and re-formation in the ISM.

\end{abstract}

\keywords{ISM: clouds --- ISM: evolution --- methods: laboratory: solid state --- molecular processes --- solid state: refractory --- ultraviolet: ISM}



\section{Introduction}

Cosmic dust grains are formed in the shells around asymptotic giant branch stars and supernovae. Stellar winds and shock waves distribute the pristine ``stardust" from the stellar environment into the interstellar medium (ISM) where it is subjected to chemical and physical processing due to UV and cosmic ray irradiation and destructive processes including erosion by sputtering, heating, and evaporation by shock waves. Eventually, dust grains are involved in star and planet formation processes.

Silicates and carbonaceous grains represent the two major components of cosmic dust. The life-cycle of cosmic dust grains is far from being understood and, in particular, the origin and evolution of interstellar grains is still under debate: are cosmic grains formed only in stellar outflows at high temperature or can they also be formed in the ISM at low temperature$?$ How do destructive processes in the ISM relate to the observed amount of cosmic dust$?$ Estimates predict that only a fraction of the dust mass entering the ISM survives under the destructive conditions present in this environment. Nanometer-sized grains are supposed to be completely shattered while micron-sized grains may survive (Jones et al. 1994, 1996; Bocchio et al. 2014; Slavin et al. 2015). According to recent studies, only a small fraction of stardust survives in the ISM (Zhukovska et al. 2008, Draine 2009, Zhukovska \& Henning 2013). This implies that in the ISM the cosmic dust destruction rate is faster than the production rate by stellar sources. However, observations of refractory matter in the ISM suggest that in order to maintain a steady amount of dust grains, a supplementary production mechanism takes place, i.e., ``in situ" re-formation of cosmic grains. Assuming amorphous hydrogenated carbon (HAC) to be one of the main forms of interstellar carbon dust, Serra D\'{i}az-Cano \& Jones (2008) have reached the same conclusions. HAC dust has a lifetime too short to allow the largest part of its population to survive in the ISM. Therefore, Serra D\'{i}az-Cano \& Jones (2008) propose the need for a re-formation mechanism of interstellar carbonaceous grains taking place directly in the ISM. Such a mechanism is required to balance the discrepancy between the timescale for stellar dust injection and its destruction in the ISM.

Laboratory investigations are required to shed light on such processes. Experiments shall focus on the re-formation of interstellar grains by accretion of gaseous precursors onto the surface of surviving grains, at low temperature. These processes may be important in the ISM (where the temperature of dust grains may be as low as 10 K; Herbst 2001) and such condensation processes are different from the growth of grains by coagulation of smaller grains. Condensation of interstellar grains may be fueled by atoms and molecules which are the product, on the one hand, of stellar and interstellar chemistry and, on the other hand, of the erosion and destruction of pre-existing grains. Condensation of SiO molecules at low temperature has been recently studied in the laboratory (Krasnokutski et al. 2014; Rouill\'{e} et al. 2014a). Reactions between SiO molecules were found to be barrierless and the experiments have clearly revealed the efficient formation of SiO$_{x}$ condensates at temperatures of about 10 K. In another experimental study, low temperature condensation of more complex silicates with pyroxene and olivine composition has been studied (Rouill\'{e} et al., 2014b). These experiments demonstrated an efficient silicate formation mechanism at low temperature. The final condensates are fluffy aggregates consisting of nanometer-sized primary grains, amorphous and homogeneous in structure and composition. Looking at the mid-infrared (MIR) spectral properties of the condensates, a significant coincidence was pointed out between the 10 $\mu$m band of interstellar silicates (Chiar \& Tielens 2006) and the 10 $\mu$m band of the low temperature siliceous condensates (Rouill\'{e} et al., 2014a).

With the present study, we want to extend our previous investigations on the low-temperature formation pathways of silicates (Krasnokutski et al. 2014; Rouill\'{e} et al. 2014a; Rouill\'{e} et al. 2014b) to carbonaceous grains. The gas-phase precursors for the formation of carbonaceous grains in the cold and denser phase of the ISM are most likely carbon and hydrogen atoms, small carbon and hydrocarbon molecules and clusters, and possibly PAHs. After the gas-phase precursor molecules condense on grain surfaces, they can react with each other or with the catalytic surfaces of the host grains. In addition, reactions can be triggered by UV irradiation or high-energy particles (see for instance, Orlando et al. 2005 and Sabri et al. 2015). Moreover, thermal heating in star-forming regions and shock heating by supernovae may be additional triggering mechanisms for surface reactions (see for instance Barlow 1978 and Orlando et al. 2005). The present experiments were designed to simulate carbonaceous grain formation under the low-temperature conditions of the ISM. Gas-phase carbonaceous precursors were created by laser ablation of graphite and accreted on cold substrates (T $\approx$  10 K) representing the cold surfaces of surviving dust grains in the ISM. The growth and evolution of the condensing carbonaceous precursors have been simulated under a variety of experimental scenarios: direct deposition on a cold substrate, isolation in Ar-matrix, grain surface diffusion and reactions after thermal annealing of the matrix and, to simulate the chemical-physical processing due to Vacuum$-$Ultraviolet (VUV) photons, exposure to VUV irradiation. We point out that the deposition in an inert matrix has the objective to simulate the conditions of isolation experienced by the gaseous species arriving onto the surface of interstellar grains while the diffusion triggered by annealing of the inert matrix has the objective to simulate the diffusion on the grain surface. This could also be, to some extent, a process analogue to diffusion inside ice matrices in the ISM. We think that this work represents an additional and necessary step toward a deeper comprehension of the mechanisms underlying the origin and evolution of interstellar grains.

\section{Experimental Procedure}

\begin{figure*}
   \centering
   \resizebox{16cm}{!}{\includegraphics[draft=false]{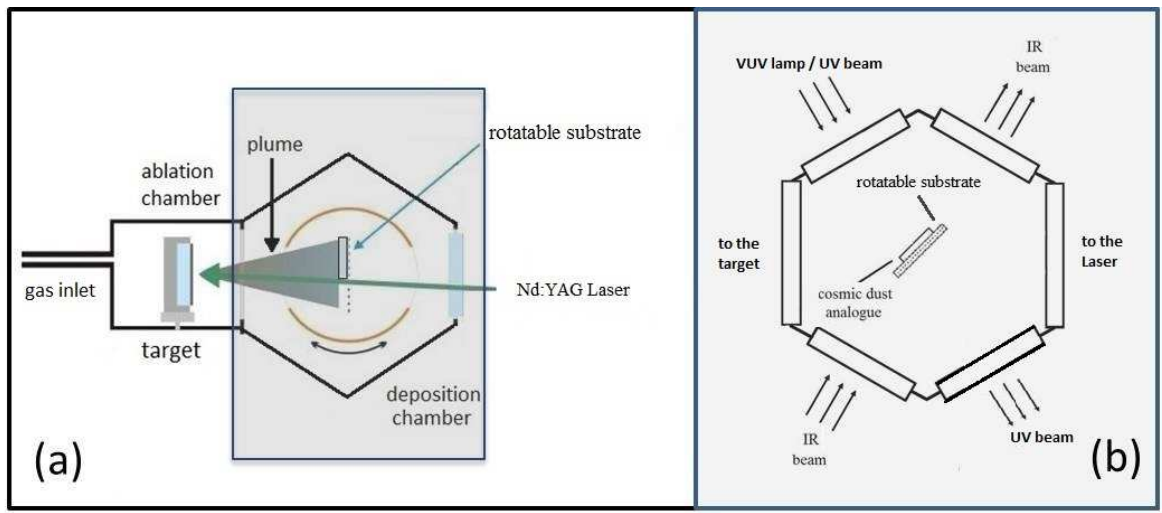}}
 \vspace{3mm}
   \caption{(a) Schematic of the laser ablation setup (ablation chamber + deposition chamber). The shadowed area delimits the deposition chamber; (b) Close-up schematic of the deposition chamber.}
         \label{fig:LaserAblationSetup}
\end{figure*}

Experiments to test the carbon condensation at low temperature were performed at the Laboratory Astrophysics Group of the Max Planck Institute for Astronomy (Heidelberg) at the Friedrich Schiller University Jena (Germany). The experimental apparatus is composed of a stainless steel high-vacuum chamber typically operating at pressure P $<$ 5$\cdot$10$^{-7}$ mbar. Graphite pellets were placed inside the vacuum chamber and used as targets for the production of the precursor species involved in the condensation of interstellar carbon analogues by laser vaporization. Sample vaporization was carried out using a pulsed Nd:YAG laser source (Continuum Minilite II) emitting photons with a wavelength of 532 nm. The laser was operated with a repetition rate of 10 pulses per second. Each pulse lasted 5 ns and carried about 10 mJ of energy. The laser beam was focused at the surface of the graphite target and, during the irradiation, the beam-spot was slightly shifted every 60 sec throughout its surface. This was done in order to vaporize a fresh part of the target after every 60 sec of irradiation and also to avoid drilling through the target. The substrate and the target for laser vaporization were separated by a distance of about 55 mm. The vaporized species were condensed onto a KBr substrate directly or isolated in solid Ar-matrix. KBr was chosen because it is a material transparent in the infrared range of interest. The substrate is placed in thermal contact with a closed-cycle helium cryostat whose temperature can go down to about 10 K. In addition, by means of a resistance heater directly connected to the cryostat, the temperature of the sample can be varied in the range 10$-$800 K. In the experiments here described, the substrate temperature was always kept between 10 and 20 K during the laser vaporization. After, the cold condensates were slowly warmed up in fixed temperature steps, typically in the range 1$-$2 K min$^{-1}$, while looking at possible changes in the MIR spectra, still under vacuum. Commercial argon (Linde, purity 99.9993$\%$) was used as matrix material. Different densities of the carbonaceous molecular species in the Ar-matrix were achieved by varying the Ar mass flow rate fed into the vacuum chamber. The Ar flow was kept between 0.75 and 1.5 standard cubic centimeters per minute (sccm). The cryostat holding the KBr substrate can be rotated, allowing the substrate to face three different directions and the corresponding pairs of opposite ports connected to the vacuum chamber. These ports allowed us to deposit the laser-vaporized species on the substrate, to acquire MIR and UV transmittance spectra, and to perform Vacuum-Ultraviolet (VUV) photo-processing experiments. Figure\,\ref{fig:LaserAblationSetup} shows a schematic of the experimental set$-$up.

For the considered samples, MIR spectra were recorded by means of a FTIR spectrometer (Bruker Vertex 80v) operated in the range of 800$-$6000 cm$^{-1}$ (1.7$-$12.5 $\mu$m), by averaging (typically) 128 scans with resolution of 1 cm$^{-1}$. The MIR beam of the FTIR spectrometer, guided along an evacuated optical path by means of gold-coated mirrors, passes through the vacuum chamber to reach an external detector. UV spectra were recorded by means of a UV spectrometer (JASCO V$-$670 EX) operated in the range of 195$-$650 nm, with resolution of 0.5 nm and a rate of 28 nm min$^{-1}$. Optical fibers fitted with collimating optics were used to carry the photons from the UV spectrometer to the vacuum chamber and back. At the beginning of each experiment, after the substrate reached the selected temperature, a background MIR spectrum and a background UV spectrum of the bare substrate were acquired. Spectra taken after each step of laser ablation (or after each step of warming up or VUV irradiation) were referred to the corresponding background spectra. We want point out that when working with transmittance spectra referred to the bare substrate, because of interference effects (coating), the resulting transmittance of the coated substrate, in a non-absorbing spectral region, can be greater than the transmittance of the uncoated substrate (i.e., the film behaves as an antireflection coating) giving a transmittance value greater than one (Macleod 1986; Palumbo et al. 2006; Fulvio et al. 2010).

During the photo-processing experiments, the vacuum chamber was interfaced to a microwave-powered hydrogen discharge lamp, commonly used to simulate and study VUV photoinduced processes on simple and complex molecules in several astrophysical environments. VUV spectra of hydrogen discharge lamps are dominated by two spectral features, centered at 122 nm and 160 nm. The lamp flux in our setup, at the sample position, is $\textit{F$_{122}$}$ $=$ 1.8 $\pm$ 0.6 $\times$ 10$^{14}$ ph cm$^{-2}$ s$^{-1}$ and $\textit{F$_{160}$}$ $=$ 3.1 $\pm$ 1.1 $\times$ 10$^{14}$ ph cm$^{-2}$ s$^{-1}$ (Fulvio et al. 2014).

The internal structure of the carbonaceous condensate was finally investigated by high-resolution transmission electron microscopy (HRTEM) after the evaporation of the ice matrix. It has been performed using a JEOL JEM 3010 microscope equipped with a LaB6 cathode operating at an acceleration voltage of 300 kV. For the sample preparation, parts of the carbon condensate was carefully scratched from the substrate and directly deposited on Cu$-$TEM grids supported by Lacey carbon films. HRTEM images were recorded digitally with a CCD camera and analyzed by the Gatan Digital Micrograph 3.9.0 software. The internal structure of the particles or condensate as a function of the experimental condensation conditions could be analyzed by performing various image analyses. The HRTEM micrographs were Fourier-Transformed (FT) to reveal any ``periodicity" in the structure. The intensity profiles of computer-generated diffractograms have been applied to visualize possible graphene layers and to distinguish between completely amorphous and more ordered carbonaceous structures. A few HRTEM images were further skeletonized by filtering the FT bright field images to eliminate non-physically related noise or periodicities. The final skeleton was compared to the original image to remove any artefacts that may have been introduced during the process. To obtain more information on the general morphology of the carbon condensate, a field emission scanning electron microscope (FESEM) (Zeiss (LEO) 1530 Gemini) was used. The resolution of the microscope depends on the material but it can be 2 nm in maximum.

\section{Precursors and formation of carbonaceous solids}

The matrix isolation technique gives us the opportunity to identify the species vaporized during laser ablation by isolating them in an inert matrix (in our case we used an Ar-matrix) deposited onto a KBr substrate. The isolation of the evaporated species in condensed rare gases cools them down before they can interact with each other. This step is necessary as laser vaporized molecules are hot and their internal energy may affect the chemistry of accretion. Moreover, the deposition in an inert matrix simulates the conditions of isolation experienced by the gaseous species arriving onto the surface of interstellar grains having temperatures between 10 and 20 K. Then, the diffusion of the isolated species can be induced by slight annealing of the matrix below the desorption temperature of the rare gas. Once the diffusion of the isolated species inside the ice takes place, they may meet each other and, eventually, react. We point out that the diffusion that we trigger by annealing of the Ar-matrix has the objective to simulate the diffusion of the precursor species on the grain surface. This could also be, to some extent, a process analogue to diffusion inside ice matrices in the ISM (keeping in mind that H$_{2}$O, CO$_{2}$ and CO are among the main constituents of interstellar ices). Of course, the annealing done in laboratory experiments has also the purpose to speed up the diffusion of species since we cannot wait for millions of years. By further warming up the condensate to a temperature higher than the desorption temperature of the rare gas matrix, any refractory material left over can be analyzed and characterized. Resuming, in matrix isolation experiments, the deposition in an inert matrix simulates the conditions of isolation experienced by the gaseous species arriving onto the surface of interstellar grains while the diffusion triggered by annealing of the inert matrix has the objective to simulate the diffusion on the grain surface and, eventually, inside ice matrices in the ISM. With this in mind, we can simulate processes in the laboratory that would happen in thousands$/$millions of years in the ISM.

\begin{figure}
   \hspace{-16mm}
   \resizebox{13.2cm}{!}{\includegraphics[draft=false]{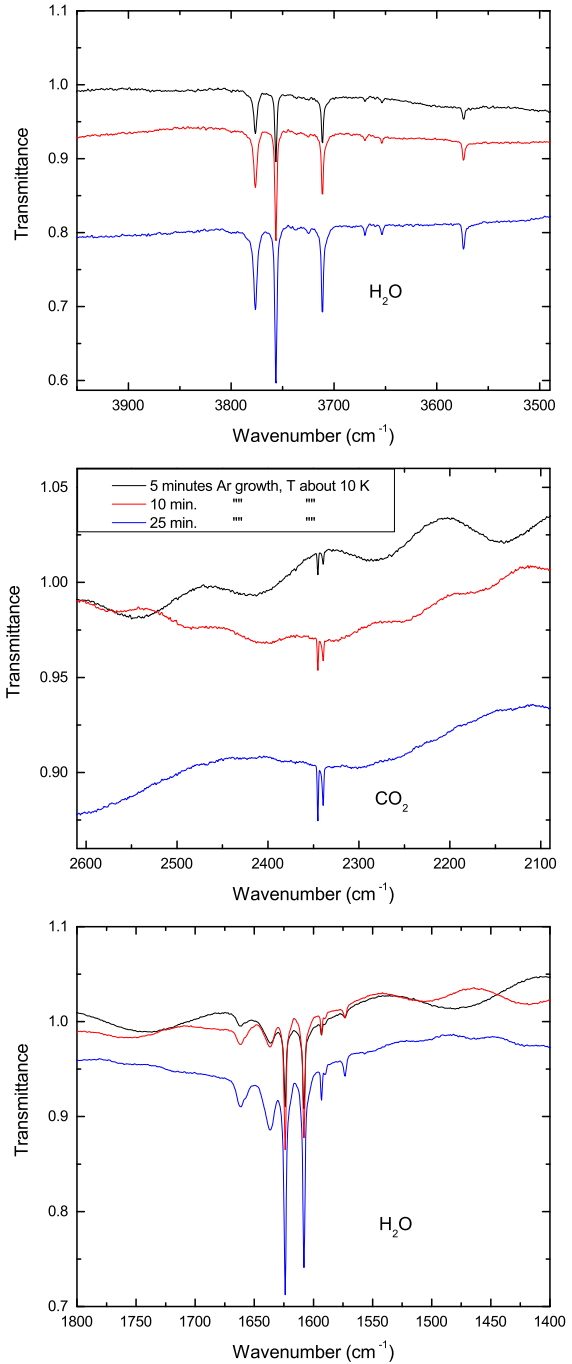}}
   \caption{Spectra acquired in a ``blank" experiment where an Ar-matrix has been condensed onto the KBr substrate at T $\approx$ 10 K, without any laser ablation. Three different spectral regions are shown. The following ``contaminant" species are seen: CO$_{2}$ and H$_{2}$O (see Table 1 and main text for details).}
         \label{fig:Blank_exp}
\end{figure}

We have verified the presence of residual gas-phase species inside the vacuum chamber (``contaminant" species). Figure\,\ref{fig:Blank_exp} shows the MIR spectra acquired in a ``blank" experiment where an Ar-matrix has been condensed onto a KBr substrate at T $\approx$  10 K, without any laser ablation. The bands present in the spectra of Figure\,\ref{fig:Blank_exp} are identified in Table 1, together with their assignment. In our experimental conditions, we can identify bands due to the following contaminant species: CO$_{2}$ (two bands around 2340 cm$^{-1}$; see for instance Schriver et al. 2000) and H$_{2}$O (set of bands around 1600 and 3700 cm$^{-1}$; Engdahl \& Nelander 1989; Perchard 2001; Hiriabashi \& Yamada 2005); the bands growth is shown at three consecutive condensation times.

\begin{deluxetable}{cccc}
\tablecaption{MIR bands and assignments of the ``contaminant" species observed in the ``blank" experiment, where an Ar-matrix was condensed on the KBr substrate at T $\approx$ 10 K, without any laser ablation of graphite.
 \label{tab:table1}}
\tablehead{
\colhead{Wavenumber (cm\textsuperscript{--1})} & \colhead{Mode} & \colhead{Molecule} & \colhead{References}}
\startdata
3776.6, 3756.3, 3737.5,                & $\nu_{3}$ & H\textsubscript{2}O & (1, 2) \\
3724.5, 3711.0                         &           &                     &        \\
3669.5, 3653.1 & $\nu_{1}$ &  H\textsubscript{2}O & (1, 2) \\
3574.1 & $\nu_{1}$ &  (H\textsubscript{2}O)\textsubscript{2} & (1, 3) \\
2345.1, 2339.3 & $\nu_{3}$ &  CO\textsubscript{2} & (4) \\
1661.4, 1636.8, 1623.8, 1607.9 &  $\nu_{2}$ &  H\textsubscript{2}O & (1, 2) \\
1592.9, 1590.1, 1573.2, 1556.3 &  $\nu_{2}$ &  (H\textsubscript{2}O)\textsubscript{2} & (1, 3) \\
\enddata
\tablecomments{\textsuperscript{1}Engdahl \& Nelander 1989; \textsuperscript{2}Perchard 2001; \textsuperscript{3}Hiriabashi \& Yamada 2005; \textsuperscript{4}Schriver et al. 2000}
\end{deluxetable}

\begin{figure}
  \hspace{-20mm}
   \resizebox{13cm}{!}{\includegraphics[draft=false]{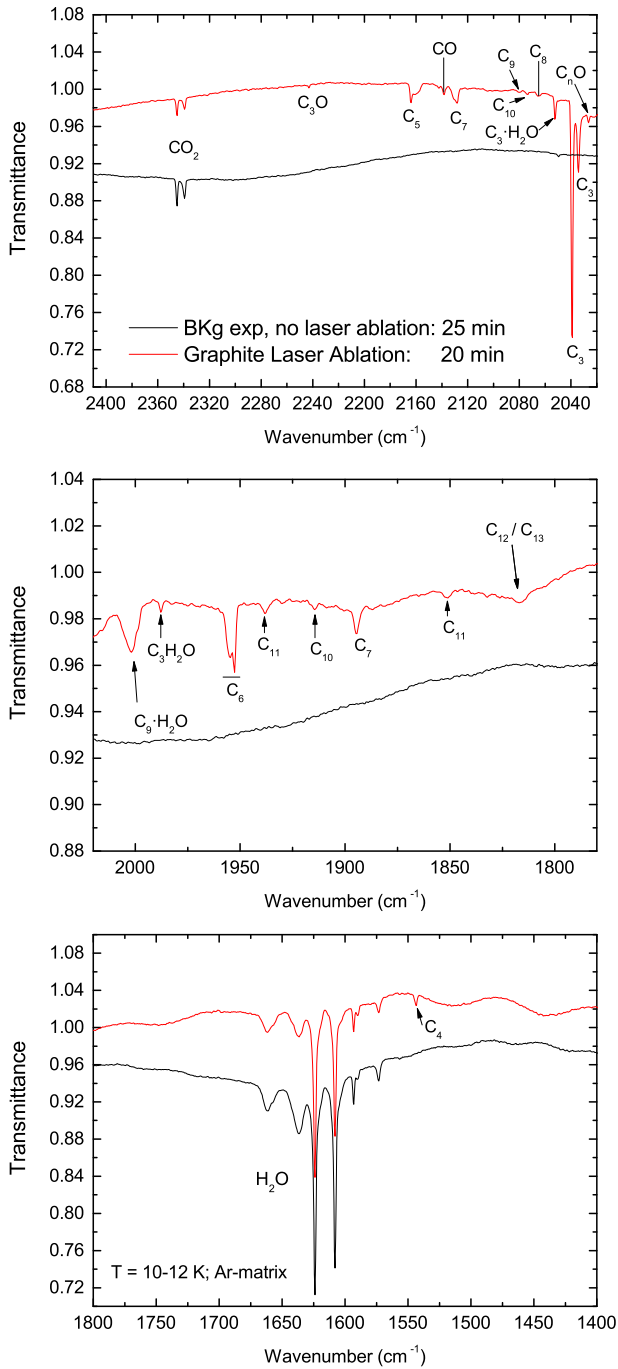}}
   \caption{Comparison between the spectra acquired after laser ablation of graphite and deposition in Ar-matrix (red lines) and a ``blank" experiment without any laser ablation (black lines). In both experiments the substrate temperature was T $\approx$ 10 K. Various carbon chains (C$_{3}$, C$_{4}$, C$_{5}$, C$_{6}$, C$_{7}$, C$_{8}$, C$_{9}$, C$_{10}$, C$_{11}$, C$_{12}$, C$_{13}$) and few oxidized species (CO and C$_{3}$O) are identified in this spectral region.}
         \label{fig:Ablation_exp}
\end{figure}

\subsection{Laser Ablation of graphite}

\begin{figure}
   \hspace{-10mm}
   \resizebox{10cm}{!}{\includegraphics[draft=false]{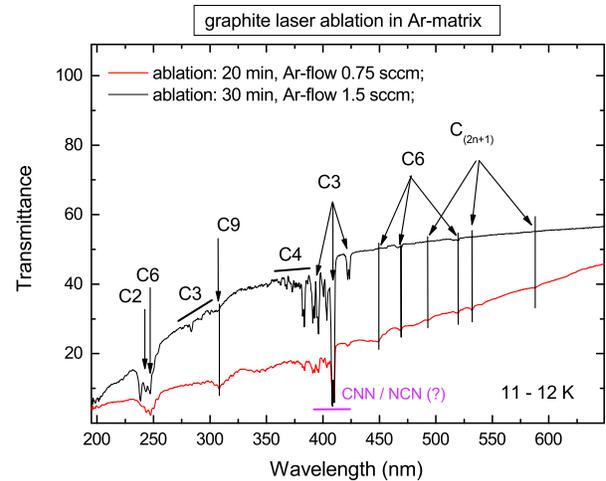}}
   \caption{UV spectra acquired after laser ablation of graphite and deposition in Ar-matrix; various carbon chains (C$_{2}$, C$_{3}$, C$_{4}$, C$_{6}$, C$_{9}$,C$_{2n+1}$) are seen in this spectral region.}
         \label{fig:Ablation_exp_UV_spectra}
\end{figure}

\begin{figure}
   \hspace{-8mm}
   \resizebox{10cm}{!}{\includegraphics[draft=false]{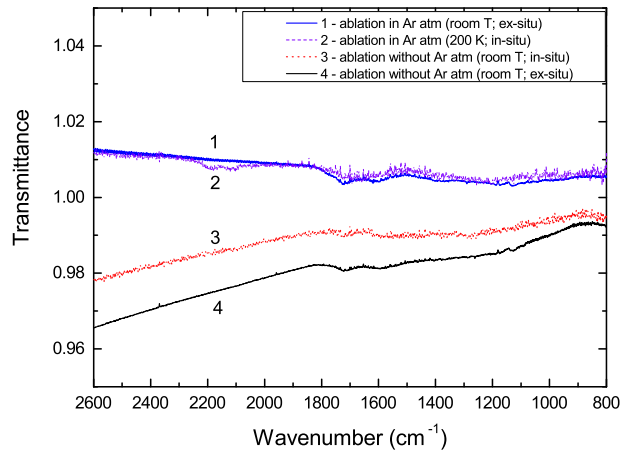}}
   \caption{Spectra of the refractory residues left over after warming up (T indicated in label) the condensates produced in the laser ablation experiments with and without Ar atmosphere; in situ: spectrum acquired under vacuum; ex-situ: spectrum acquired after exposing the sample to air.}
         \label{fig:residues}
\end{figure}

Figure\,\ref{fig:Ablation_exp} shows the comparison between the MIR spectrum acquired after laser ablation of graphite in presence of an Ar atmosphere and the spectrum of the ``blank" experiment without any laser ablation. The Ar atmosphere is used to isolate the evaporated species into a solid Ar-matrix. For the sake of brevity, we show here only selected spectral regions of main interest. The bands present in the spectra of Figure\,\ref{fig:Ablation_exp} are identified in Table 2, together with their assignment. Bands due to various carbon chains and few oxidized species (CO and C$_{3}$O) can be clearly seen besides the bands due to the CO$_{2}$ and H$_{2}$O ``contaminants" (see Table 1). Figure\,\ref{fig:Ablation_exp_UV_spectra} shows two UV spectra acquired after laser ablation of graphite in presence of an Ar atmosphere, in a complementary spectral region. The identifications of the bands present in the spectra of Figure\,\ref{fig:Ablation_exp_UV_spectra} are reported in Table 3. The presence of various isolated carbon chains is confirmed in the UV as well. The following carbon chains including C$_{2}$, C$_{3}$, C$_{4}$, C$_{5}$, C$_{6}$, C$_{7}$, C$_{8}$, C$_{9}$, C$_{10}$, C$_{11}$, C$_{12}$, C$_{13}$ were firmly identified by combining MIR and UV data. After the warming up of the condensate (i.e., the Ar-matrix plus the carbon chains and oxides in it) to room temperature, a refractory residue remained on the substrate (Figure\,\ref{fig:residues}).

\begin{deluxetable}{cccc}
\tablecaption{MIR bands and assignments of the species produced by laser ablation of graphite in presence of an Ar-matrix (see Table 1 for the identification of the ``contaminants").
 \label{tab:table2}}
\tablehead{
\colhead{Wavenumber (cm\textsuperscript{--1})} & \colhead{Mode} & \colhead{Molecule} & \colhead{References}}
\startdata
3245.2 & $\nu_{1}+\nu_{3} $ & C\textsubscript{3} & (1) \\
2242.9 & $\nu_{1}$ &  C\textsubscript{3}O & (2) \\
2163.8 & $\nu_{3}$ &  C\textsubscript{5} & (3, 4) \\
2138.2 & $\nu_{1}$ &  CO & (5) \\
2128.1 & $\nu_{4}$ &  C\textsubscript{7} & (2, 3, 6) \\
2079.9 & $\nu_{5}$ &  C\textsubscript{9} & (3) \\
2073.6 & $\nu_{6}$ &  C\textsubscript{10} & (7) \\
2065.0 & $\nu_{5}$ &  C\textsubscript{8} & (3) \\
2051.9 & $\nu_{3}$ &  C\textsubscript{3}$\cdot$H\textsubscript{2}O & (1, 8) \\
2038.9, 2034.1 & $\nu_{3}$ &  C\textsubscript{3} & (3, 7) \\
2026.4 & -- &  C\textsubscript{n}O & (2) \\
2001.8 & $\nu_{6}$ &  C\textsubscript{9}$\cdot$H\textsubscript{2}O & (8) \\
1987.5 & $\nu_{3}$ &  C\textsubscript{3}H\textsubscript{2}O & (9) \\
1954.6, 1952.6 & $\nu_{4}$ &  C\textsubscript{6} & (7, 8) \\
1938.1 & $\nu_{7}$ &  C\textsubscript{11} & (7) \\
1914.5 & $\nu_{7}$ &  C\textsubscript{10} & (7) \\
1894.3 & $\nu_{5}$ &  C\textsubscript{7} & (3, 6) \\
1850.9 & $\nu_{8}$ &  C\textsubscript{11} & (7) \\
1817.6 & $\nu_{9} / \nu_{9}$ &  C\textsubscript{12}/C\textsubscript{13} & (2, 10) \\
1543.3 & $\nu_{3}$ &  C\textsubscript{4} & (3, 11) \\
\enddata
\tablecomments{\textsuperscript{1}Ortman et al. 1990; \textsuperscript{2}Strelnikow et al. 2005; \textsuperscript{3}Freivogel et al. 1997; \textsuperscript{4}Vala et al. 1989; \textsuperscript{5}Jiang et al. 1975; \textsuperscript{6}Kranze et al. 1996; \textsuperscript{7}Cerm\'{a}k et al. 1997; \textsuperscript{8}Dibben et al. 2000; \textsuperscript{9}Szczepanski et al. 1995; \textsuperscript{10}Ding et al. 2000; \textsuperscript{11}Martin et al. 1991}
\end{deluxetable}

\begin{deluxetable}{ccc}
\tablecaption{UV bands of the species produced by laser ablation of graphite in Ar-matrix.
 \label{tab:table3}}
\tablehead{
\colhead{Wavelength (nm)} & \colhead{Molecule} & \colhead{References}}
\startdata
238.5 & C\textsubscript{2} & (1, 2) \\
243.5, 247 & C\textsubscript{2} + C\textsubscript{6} & (3) \\
275--300 & C\textsubscript{3} & (3) \\
307.5 & C\textsubscript{9} & (2) \\
362--383 &  C\textsubscript{4} & (2) \\
391--424 &  C\textsubscript{3} & (2, 4) \\
         &     CNN / NCN ($?$) &   (5)  \\
449.5, 469, 520 &  C\textsubscript{6} & (2) \\
494, 532, 588 &  C\textsubscript{n}H\textsubscript{2n+1} & (2) \\
\enddata
\tablecomments{\textsuperscript{1}Milligan et al. 1967; \textsuperscript{2}Cerm\'{a}k et al. 1997; \textsuperscript{3}Monninger et al. 2002; \textsuperscript{4}Weltner \& McLeod 1964; \textsuperscript{5}Jacox 2003.}
\end{deluxetable}

\begin{figure*}
\vspace{-10mm}
   \centering
   \resizebox{16cm}{!}{\includegraphics[draft=false]{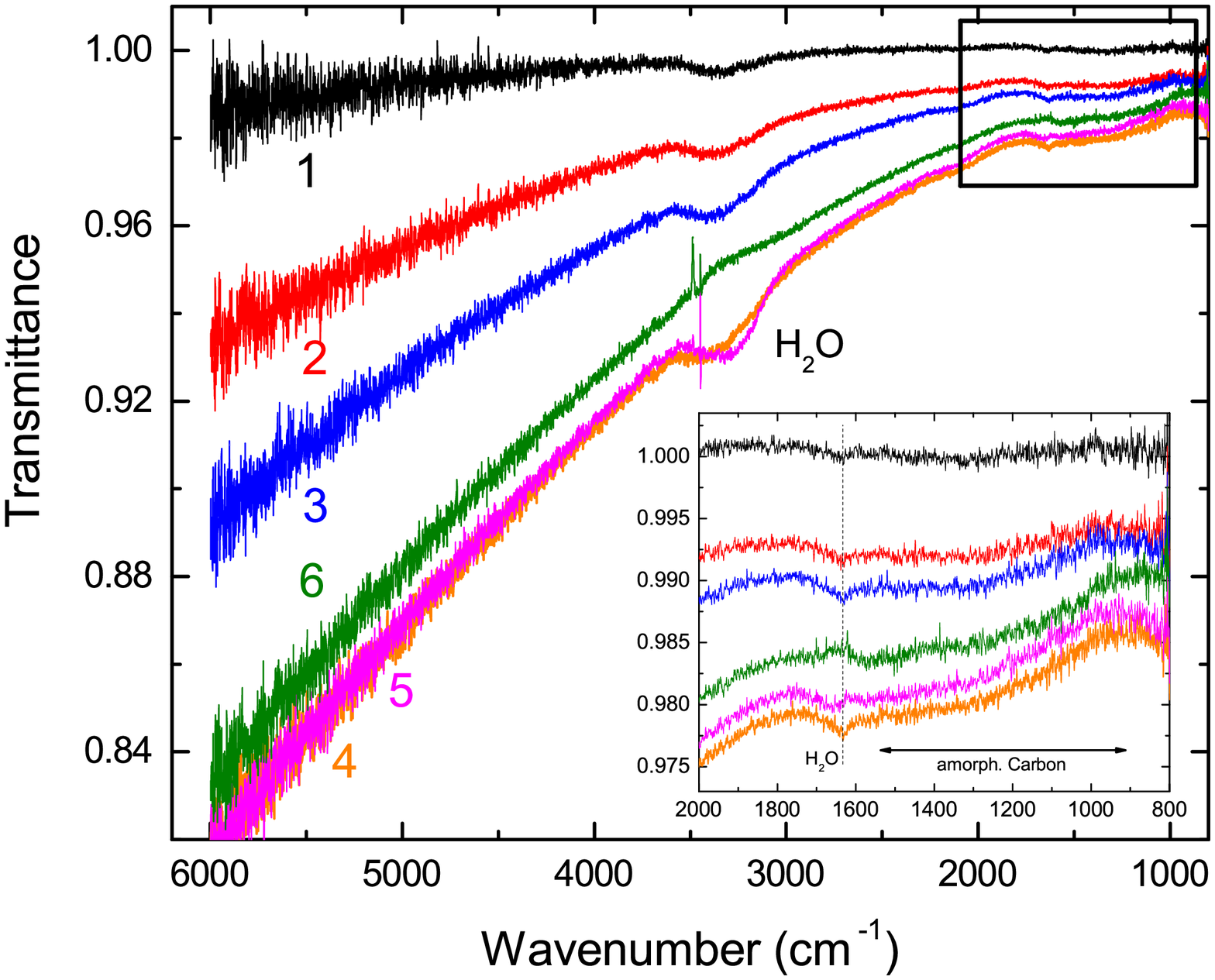}}
   \caption{MIR spectral evolution of the carbonaceous condensate formed at 10 K when laser ablation of graphite was performed without any Ar-matrix, with increasing deposition time and after warming up to 40 and 200 K, respectively. In detail, curve 1: 5 min laser ablation, T $=$ 10 K; curve 2: 15 min laser ablation, T $=$ 10 K; curve 3: 25 min laser ablation, T $=$ 10 K; curve 4: 35 min laser ablation, T $=$ 10 K; curve 5: warm up to T $=$ 40 K; curve 6: warm up to T $=$ 200 K; in the inset the spectral region between 800 and 2000 cm$^{-1}$ is magnified.}
         \label{fig:Ablation_exp_no_matrix}
\end{figure*}

As a matter of fact, the formation of a refractory residue occurs promptly when laser ablation of graphite is performed without any Ar-matrix. Infrared spectra of this case study are shown in Figure\,\ref{fig:Ablation_exp_no_matrix}. Already during the low temperature ($\approx$  10 K) phase of the experiment, a decrease of the transmission with increasing ablation time can be observed in the spectral range of 800 $-$ 6000 cm$^{-1}$. This is due to the growth of a solid refractory carbonaceous film which mainly increases the extinction in the IR range above 2000  cm$^{-1}$. The solid carbon phase that represents amorphous or hydrogenated amorphous carbon is built of plane or curved aromatic and aliphatic structural units. Free charge carriers in the aromatic areas are responsible for a continuous absorption from the UV up to the MIR and therefore, the enhanced extinction is due to continuous absorption of the carbon film and weak scattering effects. The increasing extinction is also accompanied by the arising and evolution of a broad plateau band visible between $\sim$1000 and $\sim$1400 cm$^{-1}$ (see inset in Figure\,\ref{fig:Ablation_exp_no_matrix}). The formation of a broad plateau band was often observed in amorphous and hydrogenated amorphous carbon samples prepared by gas-phase condensation techniques (J\"{a}ger et al. 1998, 1999, 2009, Schnaiter et al. 1998, Kwok et al. 2001, Llamas-Jansa et al. 2007). The weak and broad absorption plateau between 1000 and 1400 cm$^{−1}$ can be attributed to different vibrational bands of functional groups such as $\nu_{C-O-C}$, $\nu_{C-C}$, and $\delta_{C-H}$, which are finally merged into a broad MIR band in such refractory carbonaceous materials. Only little spectral changes are seen for the absorption of the refractory carbonaceous film during its warm up to 40 and 200 K, respectively. Finally, the solid refractory carbonaceous film remains on the substrate even upon warming up to room temperature (as shown in Figure\,\ref{fig:residues}).

\subsection{Laser Ablation of graphite and successive UV irradiation}

A full comprehension of the low temperature condensation processes in the ISM cannot overlook the chemical and physical processing due to VUV and cosmic ray irradiation. In this view, we have performed additional experiments where the laser ablation of graphite and deposition in Ar-matrix has been followed by VUV irradiation. The irradiation has been carried out by employing a microwave-powered hydrogen discharge lamp (see Section 2 for technical details) with an emission spectrum dominated by two spectral features, centered at 122 nm and 160 nm (Fulvio et al. 2014).

\begin{figure*}
   \centering
   \resizebox{18cm}{!}{\includegraphics[draft=false]{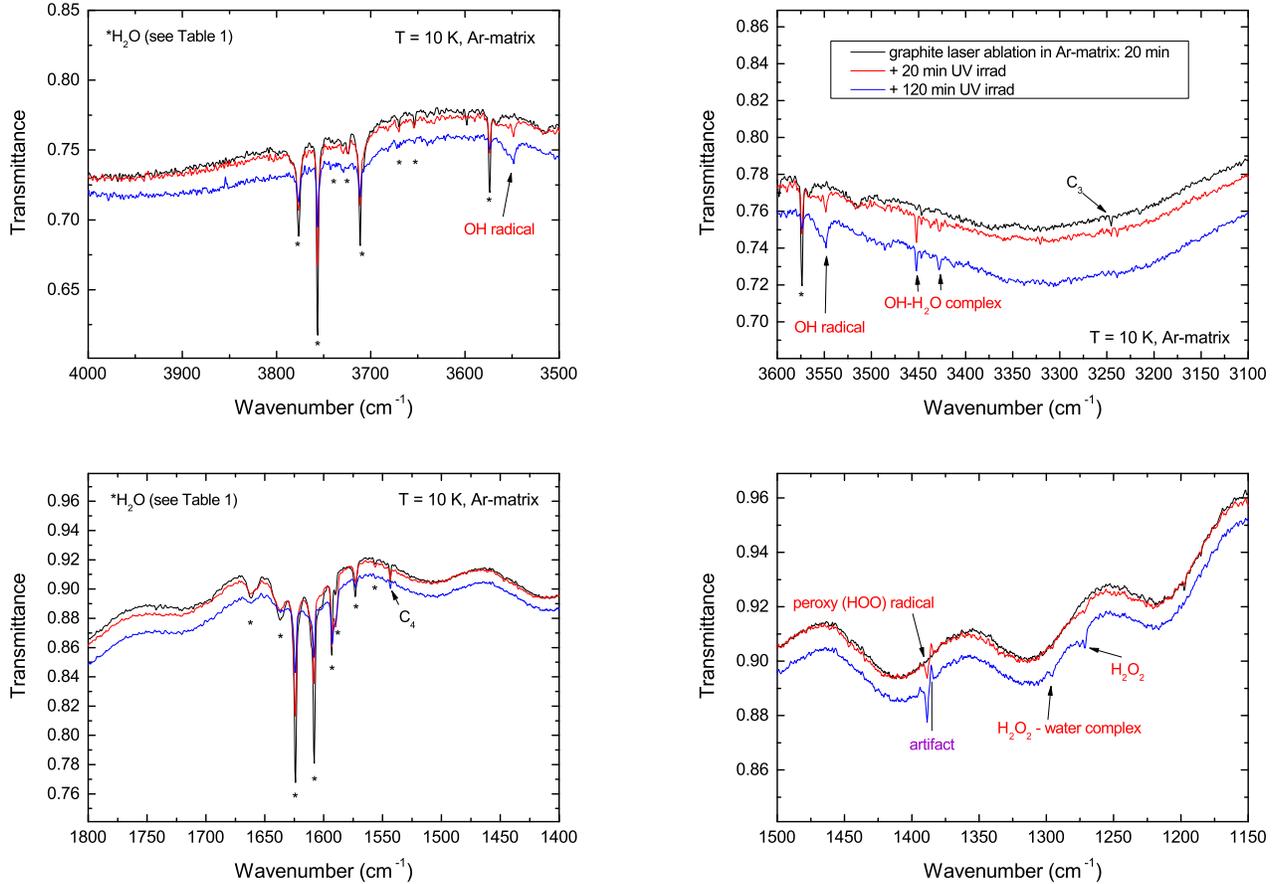}}
   \caption{Comparison between the spectra acquired after laser ablation of graphite and deposition in Ar-matrix before and after VUV irradiation (two irradiation times are shown). In the presented panels, the main spectral regions for identifying the water processing induced by the VUV irradiation are shown (new bands due to irradiation are labeled in red). The features with an asterisk refer to H$_{2}$O bands (see Table 1).}
         \label{fig:Ablation_plus_VUV_irradiation1}
\end{figure*}

\begin{figure*}
   \vspace{-20mm}
   \centering
   \resizebox{17cm}{!}{\includegraphics[draft=false]{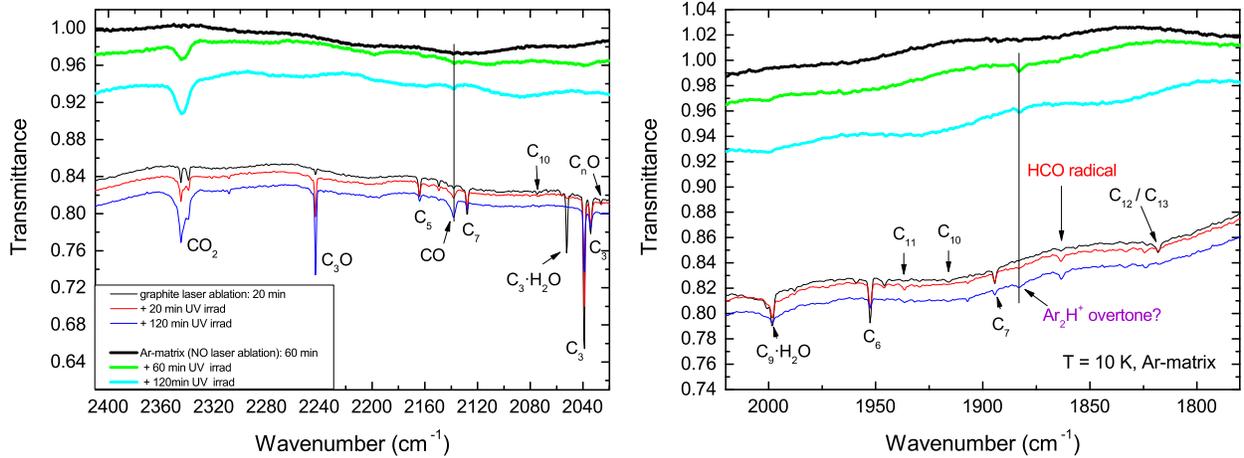}}
   \vspace{-30mm}
   \caption{Spectra acquired after laser ablation of graphite and deposition in Ar-matrix before and after VUV irradiation (two irradiation times are shown). The spectra of a ``blank" Ar-matrix deposited without laser ablation of graphite and successively irradiated with VUV photons are also shown (top side of each panel) for comparison. The main spectral regions for identifying the C-chains processing and the new molecules (labeled in red) emerging under VUV irradiation are presented.}
         \label{fig:Ablation_plus_VUV_irradiation2}
\end{figure*}

Fig.\,\ref{fig:Ablation_plus_VUV_irradiation1} and Fig.\,\ref{fig:Ablation_plus_VUV_irradiation2} show the MIR spectra acquired after laser ablation of graphite, deposition in Ar-matrix, and successive VUV irradiation. For the sake of brevity and clarity, we show only selected spectral regions of main interest and only few spectra for each experiment. In the Figures, the new bands due to the VUV irradiation are labeled in red. Their identification and assignment are provided in Table 4. All other bands seen after laser ablation of graphite and deposition in Ar-matrix without VUV processing have been already identified in Tables 1 and 2.

\begin{deluxetable}{cccc}
\tablecaption{MIR bands and assignments of the species formed under VUV irradiation in the carbon-doped Ar-matrixes.
 \label{tab:table4}}
\tablehead{
\colhead{Wavenumber (cm\textsuperscript{--1})} & \colhead{Mode} & \colhead{Molecule} & \colhead{References}}
\startdata
3548.5 & $\nu_{1}$ & OH & (1) \\
3452.6, 3428.5 & $\nu_{1}$ &  OH$\cdot$H\textsubscript{2}O & (2) \\
1863.5 & $\nu_{3}$ &  HCO & (3) \\
1388.5 & $\nu_{1}$ &  HOO & (1) \\
1295.5 &  $\nu_{6}$ &  H\textsubscript{2}O\textsubscript{2}$\cdot$H\textsubscript{2}O & (4) \\
1275.2, 1271.4 &  $\nu_{6}$ &  H\textsubscript{2}O\textsubscript{2} & (5) \\
\enddata
\tablecomments{\textsuperscript{1}Svensson et al. 2001; \textsuperscript{2}Langford et al. 2000; \textsuperscript{3}Milligan \& Jacox 1969; \textsuperscript{4}Engdahl \& Nelander 2000; \textsuperscript{5}Pettersson et al. 1997.}
\end{deluxetable}

\begin{figure}
   \hspace{-5mm}
   \resizebox{10cm}{!}{\includegraphics[draft=false]{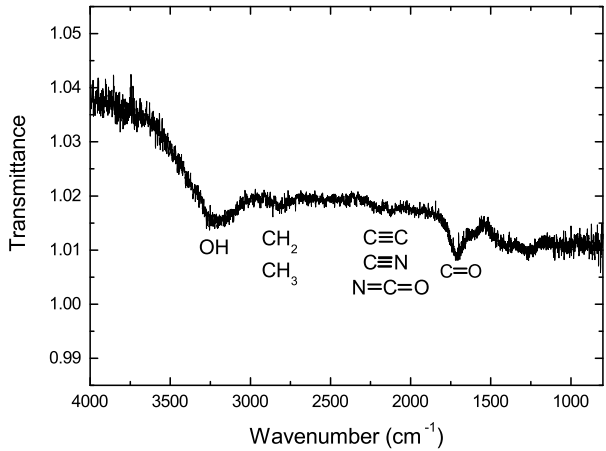}}
   \caption{Spectrum of the refractory residue left over after warming up to 200 K the condensates produced in the graphite laser ablation experiment in Ar-matrix followed by VUV irradiation.}
   \label{fig:Residue_ablation_plus_VUV_irradiation}
\end{figure}

As already discussed, the molecular species observed in non irradiated icy matrixes include various carbon chains (C$_{2}$, C$_{3}$, C$_{4}$, C$_{5}$, C$_{6}$, C$_{7}$, C$_{8}$, C$_{9}$, C$_{10}$, C$_{11}$, C$_{12}$, C$_{13}$), few oxidized species (CO and C$_{3}$O) and the ``contaminants" CO$_{2}$ and H$_{2}$O. The effects of VUV irradiation are:\\
\begin{itemize}
\item The water molecules are partially broken by the incoming VUV photons (Gerakines et al. 1996; Mason et al. 2006; Warren \& Brandt 2008) as evidenced in Fig.\,\ref{fig:Ablation_plus_VUV_irradiation1} by the continuous decreasing of all MIR water bands and by the appearing of new MIR bands related to OH, HO$_{2}$ and H$_{2}$O$_{2}$ (labeled in red). The new bands grow as irradiation proceeds as well;

\item All MIR bands related to C-chains decrease under VUV irradiation (Fig.\,\ref{fig:Ablation_plus_VUV_irradiation2}). This suggests that, like in the case of the water molecules, C-chains are partially broken by the incoming VUV photons making their carbon atoms, clusters, and radicals available for reactions. Eventually, these fragments react and form larger species such as fullerene-like structures (see Section 4);

\item The appearing and growing of MIR features related to CO, C$_{3}$O, and HCO (which indirectly confirms the previous point). We point out here that, as clearly seen in Fig.\,\ref{fig:Ablation_plus_VUV_irradiation2}, while HCO is a species purely formed under VUV irradiation, traces of CO and C$_{3}$O were already seen in the spectra before VUV irradiation. They are among the species produced by the laser ablation of graphite and deposition in Ar-matrix. Once the laser ablation stops their production stops as well. This means that the prompt growing of the CO and C$_{3}$O spectral features in the spectra showing the successive two irradiation steps is a consequence of the processing induced on the sample by VUV irradiation.
\end{itemize}

To exclude the possibility that the growing of the CO and C$_{3}$O spectral features would be related to some sort of ``contamination" rather than the VUV irradiation, we have also performed a ``blank" experiment (see Fig.\,\ref{fig:Ablation_plus_VUV_irradiation2}) where an Ar-matrix has been grown and successively exposed to VUV irradiation. We can see a clear growth of the ``contaminant" CO$_{2}$ (as expected, indeed this is a function of time) but no C$_{3}$O and only traces of CO are detectable (product of the VUV processing of CO$_{2}$).

Going back to the case study of laser ablation of graphite, deposition in Ar-matrix and successive VUV irradiation, warming up the condensate (i.e., the Ar-matrix plus the products of the VUV irradiation into it) up to 200 K, there is a refractory residue left over onto the substrate (Fig.\,\ref{fig:Residue_ablation_plus_VUV_irradiation}) as seen by the emergence of a broad plateau band visible between $\sim$1000 and $\sim$1400 cm$^{-1}$. As already discussed in the previous Section, the formation of a broad plateau band was often observed in amorphous and hydrogenated amorphous carbon samples prepared by gas-phase condensation techniques (J\"{a}ger et al. 1998, 1999, 2009, Schnaiter et al. 1998, Llamas-Jansa et al. 2007). In Fig.\,\ref{fig:Residue_ablation_plus_VUV_irradiation}, the band at about 1715 cm$^{-1}$ is due to C=O (Grishko \& Duley 2002, Llamas-Jansa et al. 2007). The other two weak bands at about 2100 $-$ 2200 cm$^{-1}$ can be due to C$\equiv$C or also C$\equiv$N (in case some N$_{2}$ was trapped into the Ar-matrix) (Grishko \& Duley 2002, J\"{a}ger et al. 2008). These groups are bound to the carbonaceous residue.

\section{FESEM and HRTEM analysis of the residues}

\begin{figure}
\centering
   \resizebox{6cm}{!}{\includegraphics[draft=false]{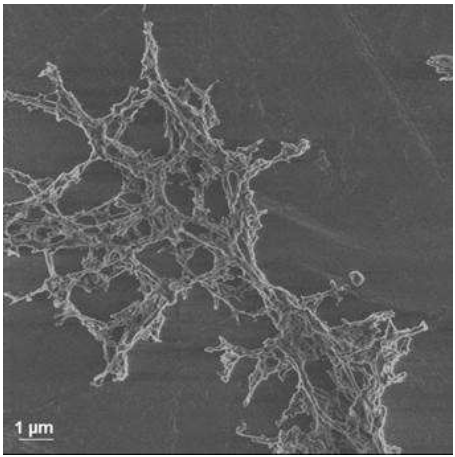}}
   \vspace{5mm}
   \caption{A large scale FESEM micrograph of the carbon condensate prepared by laser ablation of graphite and deposition of the evaporated species in an Ar-matrix.}
         \label{fig:FESEM_residue}
\end{figure}

The in situ recording of IR spectra at different temperatures and the evolution of the spectral signatures was used to prove the formation of refractory carbonaceous solids at temperatures around 10 K. To study the composition and structure of the condensate, FESEM and HRTEM were used to perform a direct imaging of the condensed carbon structures. Figure\,\ref{fig:FESEM_residue} shows a typical FESEM micrograph of a condensed layer, which is characterized by a meshed structure in a micrometer scale. The large scale structure is produced of small nanometer-sized grains forming up a porous carbon meshwork.

\begin{figure*}
\centering
   \resizebox{18cm}{!}{\includegraphics[draft=false]{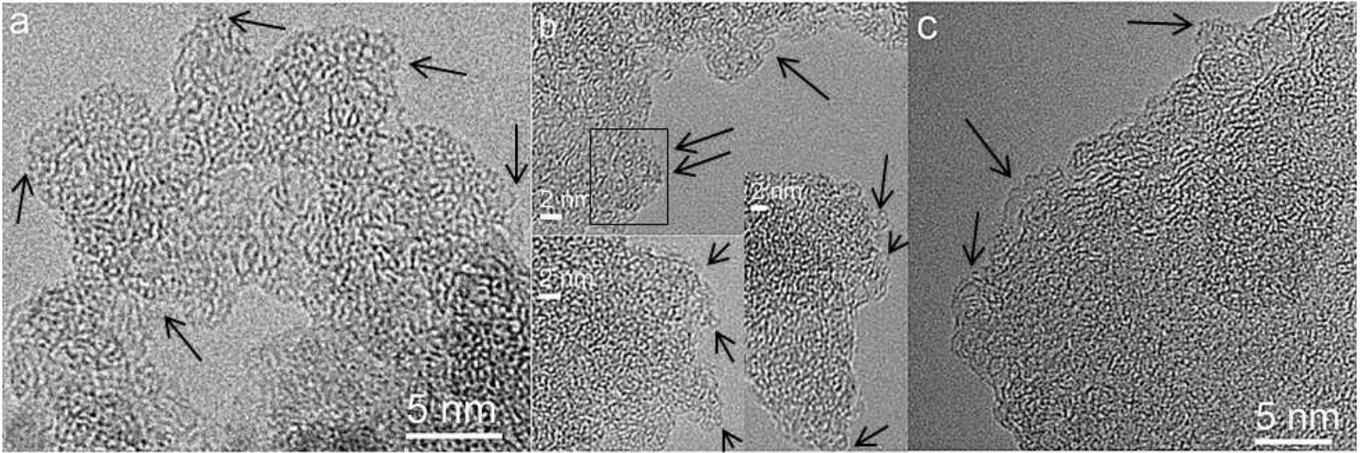}}
   \vspace{2mm}
   \caption{a) and b) HRTEM images of the condensate prepared by laser ablation of graphite and deposition in Ar-matrix. The arrows point to typical fullerene-like structures such as small bucky onions with strongly bent graphene layers and individual fullerene molecules having different sizes and shapes. The marked area in b) was skeletonized for the comparison shown in Fig. 12b; c) The carbonaceous condensate formed without Ar-matrix. In both condensates the presence of fullerenes and fullerene-like carbon structures has been proven.}
         \label{fig:HRTEM_residue}
\end{figure*}

\begin{figure}
 \vspace{3mm}
 \centering
  \vspace{2mm}
   \resizebox{8cm}{!}{\includegraphics[draft=false]{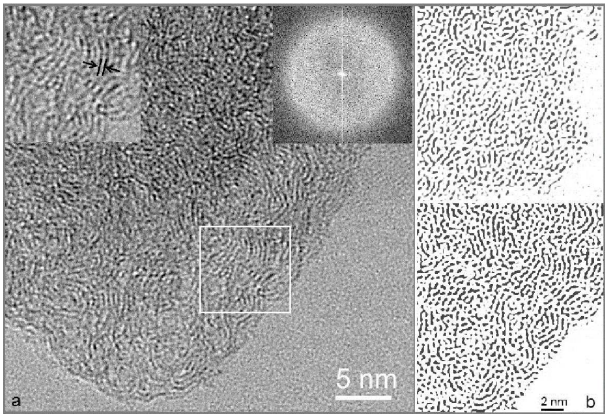}}
   \caption{a)HRTEM image of a sample condensed in an Ar-matrix plus consecutive UV irradiation. In the upper-left inset, a magnification of the marked area is presented showing an average d$_{002}$ distance of 0.391\,nm. Although the irradiated condensate appears more ordered than the non-irradiated one, the structure is still amorphous as depicted in the FT image presented in the upper-right inset; b) skeletonized micrographs produced from Figs. 11b (marked area) and Fig. 12a. The growth and stacking of the layers in the irradiated material (bottom) compared to the non-irradiated one (up) is clearly visible. See the main text for more details.}
         \label{fig:HRTEM_residue2}
\end{figure}

The analysis of the internal structure of the condensed refractory material by HRTEM reveals the presence of fullerene-like carbon material characterized by strongly curved graphene layers and the presence of various fullerenes of different sizes and shapes, which is demonstrated in the HRTEM images in Fig.\,\ref{fig:HRTEM_residue}. The term ``graphene layer" is used to describe defective and curved aromatic areas that form the 3D structure of the carbon grains. Such graphene layers may have many defects such as holes, and partly aliphatic substituents. Most of the observed graphene layers are cage fragments which are linked to each other either by aliphatic bridges or by van der Waals forces. However, also closed cage molecules (elongated or symmetric) are contained which can be detected in Fig.\,\ref{fig:HRTEM_residue}b. Only a few longer graphene layers are extending throughout the fullerene-like structures. Fullerene-like carbonaceous structures were detected in both condensates produced by deposition of the laser ablated species either on the bare cold substrate or in Ar-matrix.

The presence of differently sized cage molecules can be observed in the HRTEM images. Fullerenes are firmly detectable by HRTEM. The content of C$_{60}$ in a nanometer-sized layer of refractory carbon formed by condensation is very small and could only be detected for isolated molecules in the UV range. In addition, C$_{60}$ is not the only fullerene molecule that was formed. Smaller and larger ones are also present. The individual fullerene molecules are dispersed in the solid carbon material. C$_{60}$ and other fullerenes were frequently identified in particulate carbon samples produced in gas-phase condensation processes such as combustion or in meteorites and interplanetary dust particles (Mordkovich 2000, Rotundi et al. 2006, Becker et al. 2006). HRTEM image simulations have demonstrated that fullerene molecules produce a ring or doughnut-like structure in the HRTEM (Goel et al. 2004, Becker et al. 2006, Terrones 2010). The diameters of these rings depend on the sizes of fullerene molecules. Generally, a clear detection of individual fullerene molecules is only possible in very thin sample areas, in particular in the periphery of grains and clusters, where the superposition of different structures can be excluded. A HRTEM image simulation of superimposed amorphous carbon structures and fullerenes is rather difficult and requires detailed information on the disordered carbon structures that we do not exactly know.

The large and small scale structure of the condensed carbon is very similar to that predicted by molecular dynamics calculation, which were used to simulate the coagulation of carbon chain molecules released upon the sublimation of a rare-gas matrix containing carbon clusters (Yamaguchi \& Wakabayashi 2004). According to the authors, the condensation process is governed by carbon chains that first condense from small carbon clusters and finally convert into more stable aromatic structures. Through this, two-dimensional aromatic islands are formed linked to each other by sp$^3$ hybridized carbon structures. Eventually three-dimensional structures are formed. The conversion of sp to sp$^2$ hybridized carbon is an exothermal reaction, which may temporarily anneal adjacent areas. This release of energy may induce the growth of a few longer graphene layers within the fullerene-like carbon, which is detectable in the micrographs of the carbon condensates. The efficient cooling of the dissipated reaction energy by the cryogenic matrix hampers the growth of the long graphene layers as observed in the deposit without Ar-matrix.

The condensation process of carbon species in an Ar-matrix combined with successive irradiation by Lyman alpha photons is characterized by the initial formation of disordered fullerene-like carbon structures that evolve into more ordered structures containing larger aromatic areas under VUV processing. The higher level of ordering is clearly detectable in the HRTEM images shown in Fig.\,\ref{fig:HRTEM_residue2} compared to the original non-irradiated sample presented in Fig.\,\ref{fig:HRTEM_residue}. The evolution into more ordered structures is detectable by the increase of the length of graphene layers or aromatic areas and the decrease of interlayer distances, slowly approaching the standard value for graphite of d$_{002}$=0.335 nm. For instance, the upper-left inset in Fig.\,\ref{fig:HRTEM_residue2}a  shows a magnification of the marked area. An average d$_{002}$ distance of 0.391\,nm compared to a value beyond 0.43 nm for the non-irradiated condensate was measured. The formation of longer graphene layers with sizes between 2 and 4 nm compared to the original sample can be well noticed. However, the original fullerene-like structures are still deducible from the image. Even though clear evidence for the evolution into more ordered structures is detected, the formed material is not yet comparable to a graphitic sample. Fourier-Transform (FT) images produced of the corresponding bright field images of the irradiated carbon do not show diffraction patterns of ordered graphitic or graphene areas. As depicted in the upper-right inset of Fig.\,\ref{fig:HRTEM_residue2}a, a halo due to amorphous, disordered material is characteristic of the material. The skeletonized images in Fig.\,\ref{fig:HRTEM_residue2}b were produced from Fig.\,\ref{fig:HRTEM_residue}b (upper part) and Fig.\,\ref{fig:HRTEM_residue2}a, after removing noise by masking the FT images, which results in a higher contrast for the graphene layers. The growth and stacking of the layers compared to the non-irradiated material is clearly visible.

The condensation of carbon species at low temperatures provide similar fullerene-like carbon grains as produced in high-temperature condensation pathways of solid carbonaceous materials (J\"{a}ger et al. 2009).   Chain-like structures were also predicted and found to be precursors and intermediates in the particle formation process. Eventually, they converted into cage-like structures including fragments and completely closed cages. The role of carbon chains as precursors for the formation of fullerenes in the hot circumstellar environments of carbon-rich stars, such as Wolf Rayet stars, was also verified by theoretical condensation models (Cherchneff et al. 2000). In this scenario, polycyclic aromatic hydrocarbons are ruled out as possible intermediates. This scenario was confirmed by the experimental results presented by J\"{a}ger et al. (2009).

\section{Astrophysical Implication and Conclusions}

Laser ablation of graphite targets and deposition of the evaporated species either into Ar matrices or on bare 10 K cold substrates were used to simulate the re-formation processes of refractory carbonaceous material under conditions prevailing in the interstellar medium (ISM). The spectroscopy of the isolated species enabled us to identify relevant precursor molecules produced in the evaporation process and by reactions with ``contaminants" in the rare gas matrix. We would like to stress that the contaminants observed in our high vacuum system (molecules such as H$_{2}$O and CO$_{2}$) are normal constituents of interstellar ices, therefore they are not really a problem in the view of the conclusions drawn by our experiments and, actually, provide a more realistic astrophysical scenario (see the following discussion about new species formed under VUV irradiation).

The experiments discussed in Section 3.1 indicate that the condensation of carbonaceous solids happens already at low temperature (T as low as 10 K). This is evident when the laser ablation of graphite is performed without any isolating matrix and it is confirmed by the ablation experiments in Ar-matrix, in which the annealing of the isolating matrix is just enough to enable the isolated carbon molecules, clusters, and radicals produced by the laser ablation process to diffuse and react resulting in clusters, oligomers and, finally, refractory carbonaceous solids. The experiments also suggest that the formation process of carbonaceous solids is barrierless: neither additional energy nor long time scales are necessary to produce the solid carbonaceous materials. Interestingly, these results coincide with the finding of a barrierless condensation of silicates at temperatures of about 10 K (Krasnokutski et al. 2014, Rouill\'{e} et al. 2014a).

The experiments reported in Section 3.1 of the present work simulate what happens when gaseous carbonaceous precursors arrive onto the surface of interstellar grains, where the dust temperature can be as low as 10 K. Without any other active process, they would react with each other or with the host grains to form solid layers of carbonaceous refractory material. However, a full comprehension on the low temperature condensation processes ongoing in the ISM cannot overlook the chemical and physical processing due to VUV and cosmic rays irradiation. In this view, we have performed additional experiments (reported in Section 3.2) where the laser ablation of graphite and condensation in Ar-matrix has been followed by VUV irradiation, simulating the VUV radiation field present in various ISM regions. The irradiation has been performed by means of a microwave$-$powered hydrogen discharge lamp, commonly used to simulate VUV photoinduced processing of simple and complex molecules of astrophysical interest.

The VUV irradiation experiments reported in Section 3.2 suggest that the C-chains are partly degraded by the incoming VUV photons making their carbon atoms, clusters, and radicals available for reactions. Eventually, these fragments react and form larger species, such as fullerene-like structures or react with the water molecules present in the condensate (also broken by the incoming VUV photons). The following new species are formed under VUV irradiation: CO, C$_{3}$O, and HCO (by the reactions between the fragments of C-chains and those of water molecules) as well as OH, HO$_{2}$ and H$_{2}$O$_{2}$ (by the reactions among the fragments of water molecules with themselves). As said above, we want to stress that H$_{2}$O and CO$_{2}$ are not really a problem in the view of the conclusions drawn by our experiments. Indeed, H$_{2}$O and CO$_{2}$ are among the most abundant and important icy species in molecular clouds. The experiments in the current work have shown that the carbonaceous precursors arriving on the surface of molecular clouds grains probably react with H$_{2}$O, CO$_{2}$ and other species such as CO. The VUV photons partly dissociate these molecules and new species such as C$_{3}$O and HCO can finally be formed. By warming up the condensate (i.e., the Ar-matrix plus the products of the laser ablation and VUV irradiation), diffusion processes are triggered and the carbonaceous molecules can react, form oligomers and, eventually, carbonaceous refractory residues. Even in this case, the experiments suggest that the formation process of the carbonaceous solids is barrierless. One of the amazing findings of this study is the observation of fullerene-like carbon structures and the formation of fullerene molecules of different sizes and shapes in the HRTEM images of the refractory carbonaceous material. This indicates a formation pathway via long and branched carbon chain molecules, which are finally convert into aromatic structures as already predicted and observed for high temperature condensation processes (Cherchneff et al. 2000, Yamaguchi \& Wakabayashi 2004, J\"{a}ger et al. 2009).

As said in the Introduction, the life-cycle of cosmic dust grains is far from being understood and the origin and evolution of interstellar grains is still under debate. In particular, the following open questions need to be answered: can cosmic grains be formed directly in the ISM by means of ``in situ" re-formation mechanisms happening at low temperature$?$ How do the destructive processes in the ISM relate to the observed amount of cosmic dust$?$ Several studies have predicted that, in the ISM, the cosmic dust destruction rate is faster than the production rate by stellar sources (Serra D\'{i}az-Cano \& Jones 2008; Zhukovska et al. 2008; Draine 2009; Zhukovska \& Henning 2013; Bocchio et al. 2014; Slavin et al. 2015). However, observations of refractory matter in the ISM suggest that in order to maintain a steady amount of cosmic grains some supplementary production mechanism takes place, i.e., re-formation of cosmic grains, directly in the ISM.

We have shown in the present work that, in the ISM, low temperature condensation of carbonaceous cosmic grains is possible. The condensing gaseous precursor species are the product, on the one hand, of stellar and interstellar chemistry and, on the other hand, of the very same erosion and destruction of pre$-$existing cosmic grains. Once on the surface of cold cosmic grains: (a) without any other active process the precursor species can react with each other or with the host grains to form carbonaceous refractory aggregates and grains; (b) in presence of VUV radiation field the precursor species can be broken by the incoming photons making their carbon atoms, clusters, and radicals available for reactions with themselves or with the fragments of other molecules present on the grain surface; some of the products of the VUV irradiation will react among each other or with the host grains, and will form carbonaceous refractory aggregates and bigger grains.

In the view of the results shown in this work, the cold condensation mechanisms can give fundamental clues to fully understand the balance between the timescale for dust injection, destruction and re-formation in the ISM. Our results should be considered among the first works dedicated to simulate possible re-formation mechanisms of interstellar cosmic grains taking place at low temperature directly in the ISM.

\acknowledgments
The project was supported by the Deutsche Forschungsgemeinschaft through project No. He 1935/26-1 and JA 2102/2-2 of the Priority Program 1573 ``Physics of the Insterstellar Medium".
D. F. is also grateful to the Brazilian agency CNPq (Bolsa de Produtividade em Pesquisa - PQ 2015; Processo: 309964/2015-6). \'{A}. K. acknowledges the COST TD1308 project's support.

%





\end{document}